\documentclass[12pt]{article}
\usepackage{latexsym,cite}

\begin{document}

\title{High-precision estimates of critical quantities by means of 
improved Hamiltonians}

\author{
  \\
  {\small \hphantom{aaaaaa} Massimo Campostrini, Paolo Rossi, Ettore Vicari
          \hphantom{aaaaaa}}         \\[-0.2cm]
  {\small\it Dipartimento di Fisica and INFN -- Sezione di Pisa}    \\[-0.2cm]
  {\small\it Universit\`a degli Studi di Pisa}        \\[-0.2cm]
  {\small\it I-56127 Pisa, ITALY}          \\[-0.2cm]
  \\[-0.1cm]  \and
  {\small Martin Hasenbusch}              \\[-0.2cm]
  {\small\it Institut f\"ur Physik}    \\[-0.2cm]
  {\small\it Humboldt-Universit\"at zu Berlin}        \\[-0.2cm]
  {\small\it D-10115 Berlin, GERMANY}          \\[-0.2cm]
  \\[-0.1cm]  \and
  {\small Andrea Pelissetto}              \\[-0.2cm]
  {\small\it Dipartimento di Fisica and INFN -- Sezione di Roma I}    \\[-0.2cm]
  {\small\it Universit\`a degli Studi di Roma ``La Sapienza"}        \\[-0.2cm]
  {\small\it I-00185 Roma, ITALY}          \\[-0.2cm]
  {\protect\makebox[5in]{\quad}}  % To force authors' names to be written
                                  %   vertically, one above another.
                                  % (\author seems to put them side-by-side
                                  %   if there is room.)
  \\
}

\vspace{0.5cm}
\date{December 12, 2000}

\maketitle

\begin{abstract}
Three-dimensional spin models of the Ising and XY universality classes
are studied by a combination of high-temperature expansions and Monte
Carlo simulations applied to improved Hamiltonians.
The critical exponents and the critical
equation of state are determined to very high precision.
\end{abstract}
\clearpage

\newcommand{\<}{\langle}
\renewcommand{\>}{\rangle}
\newcommand{\be}{\begin{equation}}
\newcommand{\ee}{\end{equation}}
\newcommand{\bea}{\begin{eqnarray}}
\newcommand{\eea}{\end{eqnarray}}

%\ltapprox and \gtapprox produce > and < signs with twiddle underneath
\def\spose#1{\hbox to 0pt{#1\hss}}
\def\ltapprox{\mathrel{\spose{\lower 3pt\hbox{$\mathchar"218$}}
 \raise 2.0pt\hbox{$\mathchar"13C$}}}
\def\gtapprox{\mathrel{\spose{\lower 3pt\hbox{$\mathchar"218$}}
 \raise 2.0pt\hbox{$\mathchar"13E$}}}

\section{Introduction}

The notion of universality is central to the modern understanding of
critical phenomena.  It is therefore very important to compare
high-precision theoretical and experimental determinations of
universal quantities, such as critical exponents or universal amplitude ratios,
for systems belonging to the same universality class.

There exist several different methods for determining critical quantities.
One may study lattice models by means of high-temperature (HT) 
expansions or Monte Carlo (MC) simulations, or may consider continuum 
models and apply the well-known methods of perturbative 
field theory. All these methods require some extrapolation: 
in HT studies one wishes to determine the behavior at the critical 
point $\beta_c$ from a perturbative series around $\beta=0$; in MC studies 
that use finite-size scaling methods, an extrapolation $L\to\infty$ 
is needed; in perturbative field-theory calculations 
one must compute the value for $g=g^*$, $g^*$ being the fixed point, 
from a perturbative series in powers of $g$. The accuracy of these 
extrapolations depends on the analytic behavior of the considered functions 
at $\beta_c$ or $g^*$, which, in turn, is determined by the 
renormalization-group (RG) theory. The complex structure of the 
critical behavior, characterized by a multitude of subleading 
exponents $\omega_i$, gives rise to nonanalyticities at 
the critical (or fixed) point that make the extrapolations difficult and
often introduce large and dangerously undetectable systematic errors. 
We will not review here the field-theoretical case and we address the reader 
to Refs. \cite{Nickel-82,PV-98,CCCPV-00,CPV-00} and 
we will only discuss HT and MC methods.

The precision of the results which can be extracted from the analysis of HT
series and from MC simulations 
is mainly limited by the presence of confluent corrections with
noninteger exponents.  Let us consider, e.g., the magnetic
susceptibility $\chi(L,T)$ in a finite box $L^d$, as a function of the 
reduced temperature $t$. The analyses of HT expansions aim to determine 
the infinite-volume asymptotic behavior for $t\to 0$, which is 
\begin{eqnarray}
\chi(\infty,t) &=& C t^{-\gamma} \bigl( 1 + a_{0,1} t + ...
+\, a_{1,1} t^\Delta + a_{1,2} t^{2\Delta} + ... 
+ a_{2,1} t^{\Delta_2} + ... \bigr),
\end{eqnarray}
where $\Delta$, $\Delta_2$, ... are universal exponents. Analogously, 
MC simulations may try to determine the volume dependence 
of $\chi(L,t)$ at the critical point (or, in more sophisticated and 
efficient approaches, for a sequence of temperatures 
$t(L)$ approaching $t=0$ as $L\to\infty$):
\begin{eqnarray}
\chi(L,0) &=& \hat{C} L^{\gamma/\nu} \bigl( 1 
+\, \hat{a}_{1,1} L^{-\Delta/\nu} + \hat{a}_{1,2} L^{-2\Delta/\nu} + ... 
+ \hat{a}_{2,1} L^{-\Delta_2/\nu} + ... \bigr).
\end{eqnarray}
In both calculations the corrections with exponents $\Delta$ or $\Delta/\nu$ 
are one of the most important sources of systematic errors.
To overcome these problems one may use improved models, that is 
models for which the leading correction to scaling vanishes, i.e.
$a_{1,1} = \hat{a}_{1,1} = 0$. 
Such models can be determined by considering a one-parameter family 
of theories belonging to the given universality class, depending, say, on 
$\lambda$, and by tuning the 
irrelevant parameter $\lambda$ to the special value $\lambda^\star$
for which $a_{1,1} = \hat{a}_{1,1} = 0$; we will call such models ``improved''.

MC algorithms and finite-size scaling techniques are
very effective in the determination of $\lambda^\star$ and $\beta_c$,
but not as effective in the computation of critical exponents or other
universal quantities.  On the other hand, the analysis of HT series is
very effective in computing universal quantities, but not in computing
$\lambda^\star$ and $\beta_c$.

The strength of the two methods can be combined by computing
$\lambda^\star$ and $\beta_c$ by MC, and feeding the resulting values
into the analysis of HT series (by ``biasing'' the analysis); this
greatly improves the quality of the results.

In order to keep systematic errors under control, one may consider
several different families of models in the same universality class
and check that they give compatible results for universal quantities.

\section{Critical exponents}

Without further discussion, we present in Table~\ref{table:1} a
selection of results for the critical exponents $\gamma$, $\nu$, and
$\eta$ of the three-dimensional Ising model; for other exponents, see
Ref.\ \cite{IHT-PRE}.  IHT denotes the  results of Ref. \cite{IHT-PRE}, 
where three different improved models were considered. 
IHT${}^\star$ is a new determination in which we bias the analyses by using 
the MC estimate of $\beta_c$, as we did in Ref. \cite{IHT-CHPRV} 
for the XY model. HT is a
``traditional'' HT determination\cite{BC-00} obtained by 
analyzing the 25th-order series 
for the Ising model obtained in Ref. \cite{LCE} by means 
of biased approximants; MC are Monte Carlo
results for the $\phi^4$ improved Hamiltonians\cite{Hasenb-99} 
(see also Ref. \cite{HPV-99}); 
FT are results from the field-theoretical expansion in fixed
dimension: results (a) are taken from Ref. \cite{GZ-98}, 
results (b) from Ref. \cite{JK-99}. Other results can be found 
in Ref. \cite{PV-00}. The agreement among
the different determinations is overall satisfactory, 
although small systematic deviations are observed between the lattice 
(IHT and MC) results and the field-theoretic estimates. 
We suspect that the error estimates of Ref. \cite{JK-99} are quite too 
optimistic.

\begin{table}[t]
\label{table:1}
\setlength\tabcolsep{3.8pt}
\begin{center}
\begin{tabular}{|l|r@{}lr@{}lr@{}l|r@{}lr@{}lr@{}l|}
\hline
\multicolumn{1}{|c}{} &
\multicolumn{6}{|c|}{Ising} & \multicolumn{6}{|c|}{XY} \\
\multicolumn{1}{|c|}{}&
\multicolumn{2}{c}{$\gamma$}&
\multicolumn{2}{c}{$\nu$}&
\multicolumn{2}{c|}{$\eta$} &
\multicolumn{2}{c}{$\gamma$}&
\multicolumn{2}{c}{$\eta$}&
\multicolumn{2}{c|}{$\alpha$} \\
\hline
IHT${}^\star$ & 
      $1$&$.2372(3) $ &  $0$&$.6301(2)$ & $0$&$.0364(4) $ &
       $1$&$.3177(5)  $ & $0$&$.0380(4)$  & $-0$&$.0146(8)  $ \\
IHT & $1$&$.2371(4) $ &  $0$&$.6300(2)$ & $0$&$.0364(4) $ &
      $1$&$.3179(11) $ & $0$&$.0381(3)$  & $-0$&$.0150(17) $ \\
HT  & $1$&$.2375(6) $ &  $0$&$.6302(4)  $ &    &   &
      $1$&$.322(3)   $ & $0$&$.039(7)$   & $-0$&$.022(6)   $ \\
MC  & $1$&$.2367(11)$ &  $0$&$.6296(7)  $ & $0$&$.0358(9) $ &
      $1$&$.3177(10) $ & $0$&$.0380(5)$  & $-0$&$.0148(15) $ \\
FT(a) & $1$&$.2396(13)$ &  $0$&$.6304(13) $ & $0$&$.0335(25) $ &
      $1$&$.3169(20) $ & $0$&$.0354(25)$ & $-0$&$.011(4)   $ \\
FT(b)  & $1$&$.2403(8)$  &  $0$&$.6303(8)  $ & $0$&$.0335(6) $  &
      $1$&$.3164(8) $ & $0$&$.0349(8)$   & $-0$&$.0112(21)   $ \\
\hline             
\end{tabular}
\end{center}
\caption{Critical exponents of the
three-dimensional Ising and XY models.}
\end{table}

Similar techniques can be applied to the XY
model, with results of comparable quality.  
We present results for the critical exponents $\gamma$, $\eta$, and
$\alpha$ (we remind that $d\nu=2-\alpha$) in Table~\ref{table:1}.
They are taken from Refs.\ 
\cite{IHT-PRB1} (IHT), \cite{IHT-CHPRV} (IHT${}^\star$), \cite{BC-97}
(HT), \cite{IHT-CHPRV} (MC), \cite{GZ-98} (FT(a)), and 
\cite{JK-99} (FT(b)). Other results 
can be found in Ref. \cite{PV-00}. These 
results should be compared with the precise experimental 
estimate\cite{LSNCI-96} $\alpha =-0.01056(38)$ obtain from a Space 
Shuttle experiment for the $\lambda$
transition of ${}^4\rm He$ (cf.\
footnote 2 in Ref.\ \cite{IHT-CHPRV} for discussion of the
experimental results). There is
disagreement between IHT${}^\star$ and experiment; it would be
interesting to improve further the theoretical computation, and to
have an independent confirmation of the experimental measurement.

\section{Critical equation of state}

The equation of state relates the thermodynamical
quantities $M$, $H$, and $T$. It is easily accessible experimentally 
and thus it is important to have predictions for its behavior in the 
critical limit.

In order to determine the critical equation of state, we start from the
effective potential (Helmholtz free energy)
\be
{\cal F} (M,t) = M H - {1\over V} \log Z(H,t).
\ee
In the critical limit ${\cal F} (M,t)$ obeys a general scaling law: 
\be
\Delta{\cal F} \equiv {\cal F} (M) - {\cal F} (0) = 
   t^{d\nu} \widehat{F}_{\rm sing} (Mt^{-\beta}).
\ee
In the HT phase, ${\cal F}$ can be expanded in powers of
$M^2$ around $M=0$. We can write
\begin{eqnarray}
\Delta{\cal F} &\equiv& {\cal F} (M) - {\cal F} (0) =
{m^d\over g_4} A(z), \\
A(z) &=& {1\over2} z^2 + {1\over4!} z^4
+ \sum_{j\ge3} {1\over(2j)!} r_{2j} z^{2j},
\label{Azexp}
\end{eqnarray}
where $z\propto Mt^{-\beta}$---the normalization is fixed by  
Eq. (\ref{Azexp})---, $m$ is the second-moment mass, and 
$g_{4}$ is the
renormalized zero-momentum four-point coupling constant.  
In the critical limit, $t\to 0$, $M\to 0$ at $z$ fixed, 
the function $A(z)$ is universal. The
(universal) critical limit of $g_4$ and $r_{2j}$ can be computed from
the HT expansion of the zero-momentum $2j$-point Green's functions. For
the Ising model, we obtain\cite{IHT-PRE}
\begin{eqnarray*}
&g_4 = 23.54(4), \qquad
r_6 = 2.048(5), \\
&r_8 = 2.28(8), \qquad
r_{10} = -13(4).
\end{eqnarray*}
By using (\ref{Azexp}),
the equation of state can now be written as
\begin{eqnarray}
H(M,t) &=& {\partial{\cal F}\over\partial M} \propto
t^{\beta\delta}{d A\over d z} \equiv t^{\beta\delta} F(z),
\label{eq:H} 
\end{eqnarray}
Equivalently, one can write 
\be
H(M,t) = a M^\delta f(x),
\ee
where $x\propto t M^{-1/\beta}$ is normalized so that $x=-1$ 
corresponds to the coexistence curve and $a$ is fixed by the 
normalization condition $f(0) = 1$. The advantage of this representation 
is that by varying $x$ for $x>-1$ one obtains the full equation of state,
while, by using $F(z)$ an analytic continuation in the complex plane is 
needed to reach the coexistence curve.
The analyticity properties of $F(z)$ and $f(x)$ are constrained by
Griffiths' analyticity.

It is possible to implement all analyticity and scaling properties of
the critical equation of state introducing a parametric representation
\cite{Schofield-69,SLH-69,GZ-97} 
\begin{eqnarray*}
M &=& m_0 R^\beta \theta, \\
t &=& R(1-\theta^2), \\
H &=& h_0 R^{\beta\delta} h(\theta), 
\end{eqnarray*}
where $h(\theta)$ is normalized as $h(\theta) = \theta + O(\theta^3)$.
Note that $\theta=0$ corresponds to the HT phase $t>0, M=0$, 
$\theta=1$ to the critical isotherm $t=0$, and $\theta=\theta_0$,
where $\theta_0$ is the first positive zero of $h(\theta)$, to the 
coexistence curve.  The
analytic properties of the equation of state are reproduced if
$h(\theta)$ is analytic in the interval $[0,\theta_0)$. 
Given $h(\theta)$ the equation of state is easily obtained:
\be
f(x) = \theta^{-\delta} {h(\theta)\over h(1)}, \qquad\qquad
x = {1-\theta^2\over \theta_0^2 - 1} 
\left(\theta_0\over \theta\right)^{1/\beta}.
\ee
Approximate representations of the equation of state are obtained by 
approximating $h(\theta)$ be an odd polynomial in $\theta$, i.e. 
\be
  h(\theta) = \theta + \sum_{n=1}^k h_{2n+1} \theta^{2n+1}.
\ee
In Ref. \cite{IHT-PRE} the coefficients were determined by using IHT 
results for the constants $r_{2n}$ and a variational condition. 
More precisely, one first uses the $(k-1)$ estimates of $r_6$, ...,
$r_{2k+2}$ to fix $h_5$, ..., $h_{2k+1}$ in terms of $h_3$---in Ref. 
\cite{IHT-PRE} the variable $\rho^2 = 6 (\gamma + h_3)$ was used---
and then requires physical results to be stationary with respect to variations
of $h_3$. The idea behind this method is that in the exact case 
the function $h(\theta)$ is not uniquely defined: There exists a 
one-parameter family of equivalent $h(\theta)$. Thus, one parameter in 
$h(\theta)$ can be fixed at will. Whenever we approximate $h(\theta)$ this is 
no longer true. What we can require is that physical results have the weakest 
possible dependence on the parameter.

We use the values of $\beta$, $\delta$, $r_6$, $r_8$, $r_{10}$
obtained by IHT to compute successive approximations to $h(\theta)$;
we check the stability of the values of several universal amplitude
ratios in order to select the best approximation.  In Table~\ref{table:3}
we report the results of Ref. \cite{IHT-PRE} for the following 
amplitude ratios:
$U_0=A^+/A^-$, $U_2 = C^+/C^-$, $Q_c = B^2(f^+)^3/C^+$, $U_\xi =
f^+/f^-$. The amplitudes are defined in terms of the 
critical behavior for $t\to 0^\pm$ of the specific heat 
$C_H = A^\pm |t|^{-\alpha}$, of the second-moment correlation length
$\xi = f^\pm |t|^{-\nu}$, of the susceptibility 
$\chi = C^\pm |t|^{-\gamma}$, and of the spontaneous magnetization 
$M = B (-t)^\beta$. We compare these estimates with results 
obtained using different methods. 
HT+LT is a combination of HT and low-temperature 
expansions\cite{L-F-89,F-Z-98}; the other theoretical determinations are the
same discussed for the critical exponents, and are taken from Refs.\ 
\cite{IHT-PRE} (IHT), \cite{C-H-97,H-P-98} (MC), and
\cite{L-M-S-D-98,B-B-M-N-87,G-K-M-96} (FT).  
The agreement among
the different determinations is again satisfactory.

\begin{table}[t]
\label{table:3}
\begin{center}
\begin{tabular}{|l|llll|}
\hline
\multicolumn{1}{|c|}{ } &
\multicolumn{1}{c}{$U_0$} &
\multicolumn{1}{c}{$U_2$} &
\multicolumn{1}{c}{$Q_c$} &
\multicolumn{1}{c|}{$U_\xi$} \\
\hline
IHT  & $0.530(3)$  & $4.77(2)$  & $0.3330(10)$ & $1.961(7)$ \\
HT+LT& $0.523(9)$  & $4.95(15)$ & $0.324(6)$ & $1.96(1)$  \\
MC   & $0.560(10)$ & $4.75(3)$  & $0.328(5)$ & $1.95(2)$ \\ 
MC   & $0.550(12)$ &&& \\
FT   & $0.540(11)$ & $4.72(17)$ & $0.331(9)$ & $2.013(28)$ \\
\hline
\end{tabular}
\end{center}
\caption{Universal amplitude ratios 
for the three-dimensional Ising model.}
\end{table}

In the XY case, there are Goldstone singularities at the coexistence curve. 
In three dimensions, the leading singular behavior is correctly reproduced 
if $h(\theta) \sim (\theta-\theta_0)^2$.
We may therefore set 
\be
h(\theta) = \theta\bigl(1-\theta^2/\theta_0^2\bigr)^2
\left(1 + \sum_{n=1}^k c_n \theta^n\right).
\ee
The constant $\theta_0$ and the $k$ coefficients $c_n$ are 
fixed\cite{IHT-PRB1,IHT-CHPRV} by requiring 
the approximation to reproduce the $(k+1)$ parameters 
$r_6$, \ldots, $r_{2k+4}$.  

Only the ratio $U_0= A^+/A^-$ is measured experimentally to high 
precision.\cite{LSNCI-96}  Such a ratio has been determined in Ref. 
\cite{IHT-CHPRV} using the equation of state and the IHT${}^\star$ estimate of 
$\alpha$: They obtain 
$U_0 = 1.062(4)$. This result disagrees with the experimental estimate 
$U_0 = 1.0442$. However, the estimate of $U_0$ is strongly correlated to that 
of $\alpha$. Indeed, by using a slightly lower value of $\alpha$, 
$\alpha = -0.01285(38)$, Ref. \cite{IHT-PRB2} found $U_0 =1.055(3)$. 
Thus, the disagreement between the estimate of Ref. \cite{IHT-CHPRV} 
and experiment can be reconduced to the discrepancy in $\alpha$.
Other estimates of $\alpha$ have been obtained by using the fixed-dimension 
expansion,\cite{L-M-S-D-98} $U_0 = 1.056$, and the 
$\epsilon$-expansion,\cite{B-B-M-N-87} $U_0 = 1.029(13)$.

\section{Conclusions}

The study of HT series of ``improved'' models, with parameters
determined by MC simulations, allowed us to compute with high
precision the universal quantities---critical exponents and equation of 
state---characterizing the critical behavior of the symmetric
phase.

Suitable approximation schemes allow the reconstruction of the
critical equation of state starting from the symmetric phase; many
universal amplitude ratios can be computed.

For the Ising universality class, theoretical computations are much
more precise than experiments.  On the other hand, for the XY class,
some very precise experimental results for $\alpha$ and $A^+/A^-$ have
been obtained.\cite{LSNCI-96}  There is disagreement with the most
precise theoretical results.\cite{IHT-CHPRV}  A new-generation
experiment is in preparation \cite{Nissen-98}; it would be interesting
to improve further the theoretical computations as well.

\end{document}